\documentclass[12pt]{iopart}
\usepackage{graphicx}
\begin{document}
\title[Enhanced transmission of slit arrays]{Enhanced transmission of 
slit arrays in an extremely thin metallic film}
\author{A. Moreau, C. Lafarge, N. Laurent, K. Edee and G. Granet}
\address{LASMEA, UMR CNRS 6602, Universit\'e Blaise Pascal, 24 avenue
  des Landais, 63177 Aubi\`ere, France.}

\begin{abstract}
Horizontal resonances of slit arrays are studied. They can lead to an enhanced
transmission that cannot be explained using the single-mode
approximation. A new type of cavity resonance is found when the slits
are narrow for a wavelength very close to the period. It can be
excited for very low thicknesses. Optimization shows these structures
could constitute interesting monochromatic filters.
\end{abstract}
\submitto{\JOA}
\maketitle

Since the discovery of Ebbesen that subwavelength hole arrays could
transmit light,\cite{ebbesen} much has been understood concerning the
behavior of such structures. Resonances of the structure are
responsible for the extraordinary transmission. Two types of
resonances are usually involved : surface resonances (surface
plasmons, often refered to as horizontal resonances) and
cavity resonances (or vertical resonances).

Slit arrays (in the case of TM polarization) first attracted much attention, for they were
considered simpler than hole arrays. Porto {\it et
  al} were the first to introduce the single-mode approximation which
allows to understand the enhanced transmission\cite{porto}. They came to the
conclusion that both surface plasmon resonances and cavity resonances
were responsible for the enhanced transmission. These structures were
then much been studied on this basis\cite{garcia-vidal,collin,popov,lalanne1}. The
controversy began when Cao and Lalanne expressed the idea that
the surface plasmons could even hinder the transmission\cite{cao}.
The controversy is not closed yet\cite{moreno,fplike,lalanne3}.

Finally, hole arrays appear to be much better understood
now than slit arrays. Surface plasmons are responsible for the transmission in the case
of circular holes\cite{popov,bonod,lalanne4}. For coaxial hole arrays, the
excitation of cavity modes explains the enhanced transmission\cite{vanlabeke,moreau}.

This paper deals with the case of subwavelength slit arrays in the
visible domain. Many previous works concern the infra-red 
domain\cite{porto,garcia-vidal,lalanne1}. The permittivity of
silver and gold in this domain is very high, so that the wavelength at
which the surface plasmons can be expected are very close to the
apparition of a new diffraction order. It is difficult in this case
to distinguish between surface resonances and Rayleigh anomalies.

Our purpose is to have a sound discussion about the different modes of
the structure, using previous results on this
subject\cite{garcia-vidal,collin,lalanne3}.
The single-mode approximation and the limits of its validity have a central
role in the discussion. In the following, we show that cavity
resonances can be excited for very low thicknesses of the structure,
and that there are resonances even when the one-mode assumption is
not valid any more.

\section{The single-mode approximation}

\begin{figure}[htb]
\centerline{\includegraphics[width=8.3cm]{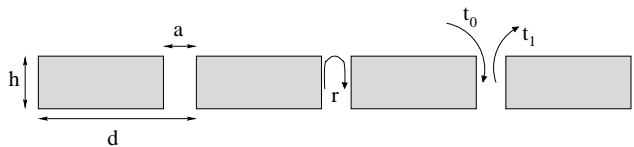}}
\caption{Geometric parameters of the slit array.}
\label{fig:dessin}
\end{figure}

Let us consider a slit array (see figure\ref{fig:dessin}), whose parameters are its thickness $h$, its
period $d$ and the width of the slits, $a$. The metal considered here is
silver. The optical constants of silver are taken from
\cite{gold}. In the following, we will consider only the case of TM
polarization, since the enhanced transmission occurs for this
polarization only.

Among the Bloch modes that are supported by a slit array,
only one of them is always propagative. It corresponds to the excitation in
each slit of the only mode which has no cut-off, in TM
polarization. The slits can be considered independent.

It has been early recognized\cite{porto} that taking this only mode into account
is enough to explain the enhanced transmission of the slit array. This
is the single-mode approximation. Some more
elaborate models have been proposed\cite{lalanne1,garcia-vidal} which rely on
more assumptions about the propagating mode or the coupling between the incident
wave and the guided mode. We will not use such approximations here.

The assumption that only the propagative Bloch mode is responsible for the
extraordinary transmission leads to a classical Fabry-P\'erot formula
for the zero-order transmittance of the whole structure \cite{porto,garcia-vidal,lalanne3} :
\begin{equation}
T=\left|\frac{t_0\,t_1\,e^{i\beta\,h}}{1-r^2\,e^{2\,i\beta\,h}}\right|^2
\label{e:fp}
\end{equation}
where $h$ is the depth of the structure, and
$\beta=\frac{2\pi}{\lambda_g}$ is the propagation constant of the
guided mode and $\lambda_g$ its effective wavelength. The coefficient
$t_0$ is the transmission coefficient between the incident wave and the
propagative Bloch  mode, while $t_1$ is the transmission coefficient between 
the propagative mode and the plane wave (see figure \ref{fig:dessin}). 

This model allows to explain the resonances of the full structure
using the properties of the semi-infinite structure - that is to say
the interface between air and the slit array. The coefficients $r$,
$t_0$ and $t_1$ can actually be found in the scattering matrix of this
interface alone, while $T$ belongs to the scattering matrix of the
whole structure with a finite thickness. The discussion will now focus
on the modes of these two structures (the semi-infinite one and the
whole structure), which correspond to poles of their respective
scattering matrixes\cite{collin}. The study of these poles has been
early recognized as central\cite{garcia-vidal} and they have recently attracted
some attention\cite{lalanne3}.

The horizontal modes\cite{collin} can be defined as the modes of the
semi-infinite structure, since they correspond to an enhanced field
at the interface between air and the slit array. As we will see, these
modes may or may not be responsible for the enhanced transmission of
the whole structure. The resonances which
appear for the full structure only and which cannot be related to any
horizontal mode will be called ``vertical resonances''.

In this work, the scattering matrix is computed using the Rigorous
Coupled Wave Analysis\cite{rcwa1,rcwa2}. It is easy to identify the
only propagative mode among the Bloch modes of the slit array, using
its propagation constant $\beta$. Then the coefficient $r$, $t_0$ and
$t_1$ are taken from the scattering matrix. As was shown in previous
works\cite{garcia-vidal,lalanne1} a full numerical approach is not
necessary for the determination of these coefficients since the
propagative mode is very close of the propagative mode in a perfect
metal waveguide. But we will not use any of these approximations here.

The one-mode assumption is very accurate\cite{garcia-vidal,lalanne1,lalanne3}. It perfectly
accounts for all the observed enhanced transmissions until now : an enhanced transmission occurs
when the denominator of (\ref{e:fp}) is close to zero. This occurs when two conditions are
fulfilled. First the modulus of $r$ should not be null, and should be as close to
$1$ as possible. Second, if we denote $\phi$ the phase
of $r$, the resonance condition can be written 
\begin{equation}\label{e:res}
\phi(\lambda)+\beta\,h=m\,\pi,
\end{equation}
where $m$ is an integer.

\begin{figure}[htb]
\centerline{\includegraphics[width=8.3cm]{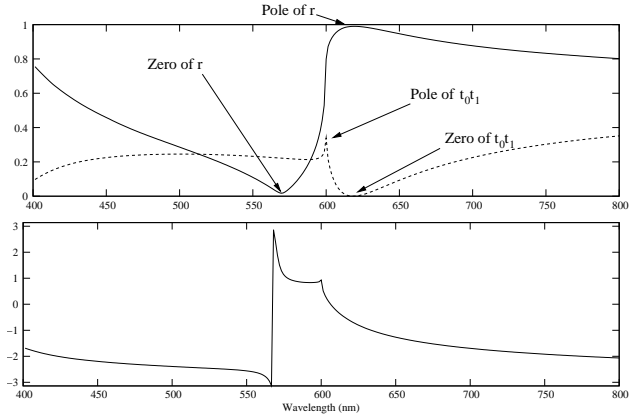}}
\caption{Typical example of $|r|$ (top, solid line), $|t_0\,t_1|$
  (top, dotted line) and $\phi$ (bottom, solid line) for the
  semi-infinite structure with $d=600\,nm$ and $a=100\,nm$.}
\label{fig2}
\end{figure}

Figure \ref{fig2} shows a typical internal reflexion coefficient
$r$ for a period $d=600\,nm$ and a slit width $a=100\,nm$ and the
modulus of $t_0\,t_1$. These quantities are enough to fully understand
the properties of the slit array.

\section{Horizontal resonances}

We will now try to identify the resonances of the semi-infinite
structure. Since it has been demonstrated that the propagative mode is
the only one responsible for the enhanced transmission, one can expect the
coefficients $r$, $t_0$ and $t_1$ to present poles for these
horizontal resonances.

Two poles can be identified in figure \ref{fig2} : one on $|r|$ and one
on $|t_0\,t_1|$.

\subsection{Pole of $r$}

The pole of $r$ has already been much
studied\cite{garcia-vidal,lalanne3}. It is located at the very wavelength where a
surface plasmon can be expected in normal incidence for $\lambda >
d$. The wavelength of the surface plasmon resonance is given
by\cite{porto,garcia-vidal,cao} 
\begin{equation}
\lambda = d\,Re\left(\sqrt{\frac{\epsilon(\lambda)}{1+\epsilon(\lambda)}}\right).\label{e:plasmon}
\end{equation}
 
Since the modulus of $r$ is close to one near this resonance, it was
first considered responsible for the enhanced transmission that could
occur in the vicinity of $\lambda_p$\cite{garcia-vidal}, strengthening
the conclusion that the horizontal resonances played an important 
role\cite{porto}. 

But,  as  underlined\cite{cao,lalanne3},
$t_0\,t_1$ presents a (real) zero  for the same wavelength so that the
transmission is  null for $\lambda=\lambda_p$.   This is the  case for
energetic reasons  : when the  reflection coefficient $r$ is  close to
$1$,   $t_1$   goes  to   zero   and   so   does  $t_0$   because   of
reciprocity. Actually $t_0$ and $t_1$ very slightly differ because the
metal is lossy, but they present the same poles and zeros.

\begin{figure}[htb]
\centerline{\includegraphics[width=8.3cm]{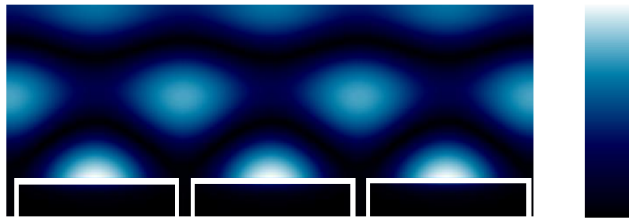}}
\caption{Modulus of $H_y$ for the semi-infinite structure with
  $d=600\,nm$, $a=10\,nm$ and $\lambda=\lambda_p=618\,nm$. A
surface resonance is clearly excited. There is
  almost no field inside the slits. The borders of the structure have
  been represented in white.}
\label{fig3}
\end{figure}

This behavior can be explained by the nature of this horizontal
resonance : it corresponds to the excitation of two counter-propagative surface
plasmons which interfere destructively above the slits so that
(i) the  propagative mode of the  slits is not
excited and (ii) the location of  the pole of $r$ (and subsequently of
the zero of $t_0\,t_1$) is  exactly $\lambda_p$ and does not depend on
the size of the slits. Figure \ref{fig3} shows the field at the
surface of the semi-infinite structure for $d=600\,nm$ and $a=10\,nm$ 
at $\lambda=\lambda_p$. The horizontal resonance can clearly be seen,
and almost no field is present in the slits. This is the case even for
higher values of $a$.

Practically, the transmission presents a zero for $\lambda=\lambda_p$,
exactly when the horizontal mode is excited so that it is difficult to
consider this resonance responsible for the enhanced transmission.

Moreover, the first works agreed that all the enhanced transmission
that have been numerically observed until now occur when (\ref{e:res})
is satisfied\cite{garcia-vidal}. All the resonances then present all the properties of
Fabry-Pérot resonances, or cavity resonances\cite{popov}. This is why we think
that they should all be labeled ``vertical resonances''\cite{collin}.

We think that the sharpness of a resonance should not be used as a
criterium to determine whether a resonance is horizontal or 
vertical\cite{porto,collin,garcia-vidal}. Otherwise this leads to apparent
paradoxes\cite{fplike}.

\subsection{Pole of $t_0\,t_1$}

Let us consider the second horiizontal resonance, that has not been
studied in previous works\cite{lalanne3} mainly because $t_0\,t_1$ has
not attracted all the attention it deserves\cite{garcia-vidal}.

\begin{figure}[htb]
\centerline{\includegraphics[width=8.3cm]{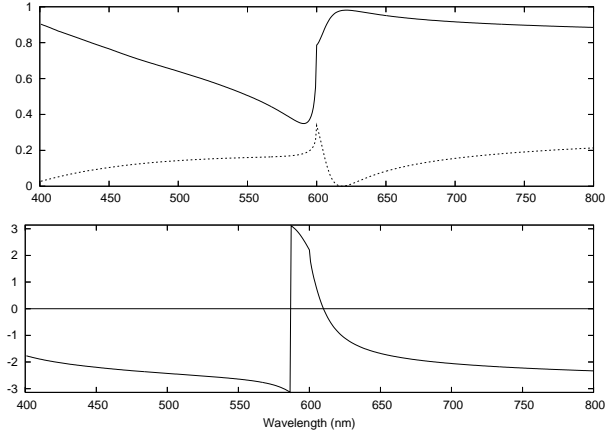}}
\caption{Modulus of of $r$ (top, solid line), $|t_0\,t_1|$
  (top, dotted line) and phase of $r$ (bottom, solid line) for the
  semi-infinite structure with $d=600\,nm$ and $a=40\,nm$. The zero of
  $r$ moves towards the left, and the quick variation of the phase
  follows. The poles of $r$ and $t_0\,t_1$ do not move.}
\label{fig4}
\end{figure}

\begin{figure}[htb]
\centerline{\includegraphics[width=8.3cm]{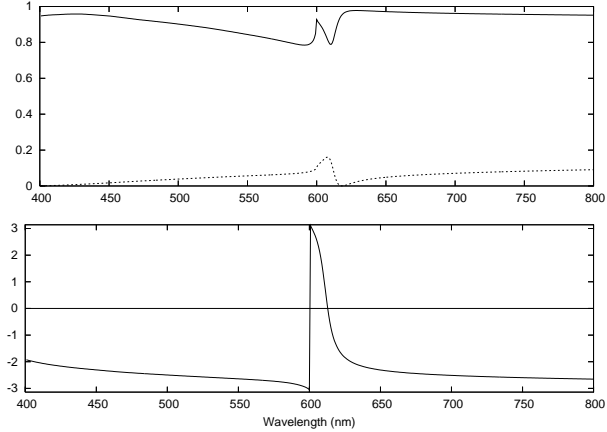}}
\caption{Modulus of of $r$ (top, solid line), $|t_0\,t_1|$
  (top, dotted line) and phase of $r$ (bottom, solid line) for the
  semi-infinite structure with $d=600\,nm$ and $a=10\,nm$. The pole of
  $t_0\,t_1$ moves towards the pole of $r$. It is located at $\lambda=608\,nm$.}
\label{fig5}
\end{figure}

First, does the observed peak really correspond to a surface resonance
? Figure \ref{fig4} and \ref{fig5} show the behavior of the usual
coefficients when $d=600\,nm$ and when $a$ becomes smaller
(respectively $a=40\,nm$ and $a=10\,nm$). For $a=40\,nm$ the zero of
$r$ is the only remarkable feature which moves : it is heading towards
$\lambda=d$, and off the real axis and the rapid variation of the phase is following this zero.
For very low values of $a$, the peak of $t_0\,t_1$ moves towards 
$\lambda_p$ and a minimum of $r$ appears at the very wavelength of the
peak ($\lambda=608\,nm$ for $a=10\,nm$). 

\begin{figure}[htb]
\centerline{\includegraphics[width=8.3cm]{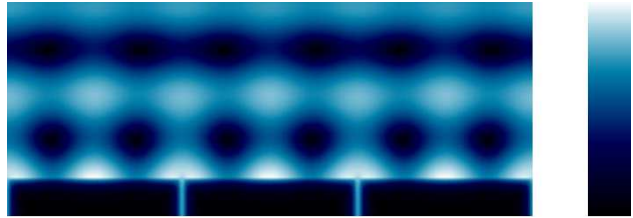}}
\caption{Modulus of $H_y$ for the semi-infinite structure with
  $d=600\,nm$, $a=10\,nm$ and $\lambda=608\,nm$. A
  surface resonance is clearly excited. It is very well coupled to the
  propagative mode inside the slits.}
\label{fig6}
\end{figure}

Figure \ref{fig6} shows the modulus of the magnetic field in the case
of the semi-infinite structure for $d=600\,nm$ and $a=10\,nm$ at the
wavelength of the peak of $t_0\,t_1$ ($\lambda=608\,nm$). It has all
the characteristics of a horizontal resonance : the field is enhanced
near the interface. This time, there is a constructive interference
of the surface modes above the slits so that (i) the propagative mode
inside the slits is easily excited and thus $t_0$ and $t_1$ present a
peak and (ii) the position of this peak strongly depends on the size
of the slits and is not located at $\lambda=\lambda_p$.

When the thickness $h$ of the slit arrays is rather large, this
horizontal resonance is responsible for a peak in the spectrum at $\lambda=d$,
which is perfectly well reproduced by the one-mode model. Absolutely
no cavity resonance is involved in this phenomenon for the denominator
of (\ref{e:fp}) does not present any noticeable minimum in this range.

\begin{figure}[htb]
\centerline{\includegraphics[width=8.3cm]{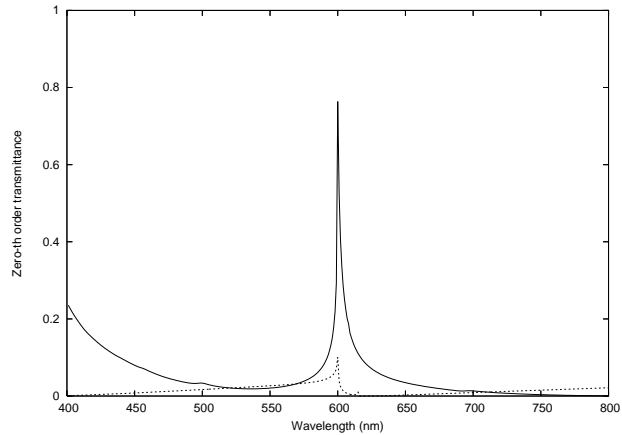}}
\caption{Zero-th order transmission of a slit array with
  $d=600\,nm$,$a=51.37\,nm$ and $h=33.89\,nm$ (solid line) and the
  single-mode approximation result (dashed line). The one-mode model
  is clearly inaccurate for these parameters : the enhanced transmission
  is due to the excitation of a horizontal resonance. The geometrical
  parameters have been obtained using optimization (see thereafter).}
\label{fig7}
\end{figure}

When the thickness of the structure becomes really small, the spectrum
shows a much more intense transmission for $\lambda=d$ (see figure \ref{fig7}). This
transmission {\em cannot be explained using the single-mode
  approximation}, since the one-mode model reproduces very poorly
the enhanced transmission. This means that not only the propagative
mode is excited by the surface resonance. Many evanescent modes
obviously are excited. While $h$ is large, they do no take part to the
transmission, but they do when $h$ is small enough. Such a resonance
cannot have been studied up to now, since all the resonances in
previous works\cite{porto,garcia-vidal,collin,lalanne3} are very
accurately described by the one-mode model. The only exception, in a
sense, is a recent and interesting work in the case of TE polarization\cite{moreno}.

\begin{figure}[htb]
\centerline{\includegraphics[width=8.3cm]{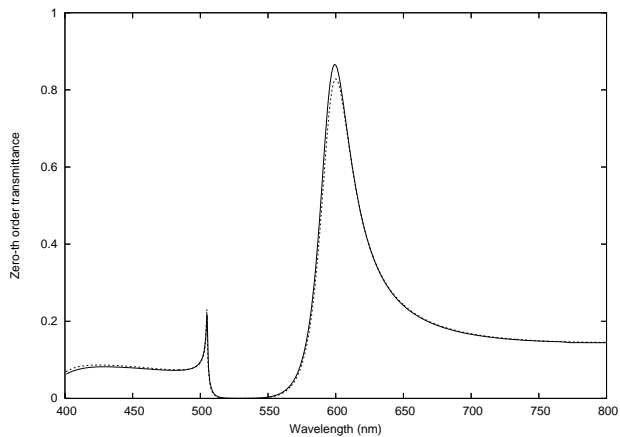}}
\caption{Zero-th order transmission of a slit array with
  $d=504.9\,nm$,$a=115.7\,nm$ and $h=127.9\,nm$ (solid line) and the
  single-mode approximation result (dashed line). The peak at
  $\lambda = 504.9\,nm$ is due to the excitation of the horizontal
  resonance on both sides of the structure. The parameters of this
  structure have been found using optimization, so that the $0$
  resonance has a maximum $@600\,nm$.}
\label{fig8}
\end{figure}

\section{Cavity resonances at very low thicknesses and optimization}

When $h$ is very small compared to the wavelength, no vertical
resonances (or cavity resonances) are expected\cite{porto,collin2,popov,lalanne3}.
This may be true when $r$ is real, but around $\lambda=d$ the phase
of $r$ takes all the values in $[-\pi,\pi]$ so that cavity resonances
should be expected for any height.

If $h$ is small, then the cavity resonances can be found either when
$m=0$ in equation (\ref{e:res}) (in that case the phase of $r$ is
negative an close to zero and the resonance will be called ``0
resonance'') or when $m=1$ (in that case the phase is
positive and close to $\pi$ and the resonance will be called ``$\pi$ resonance''). 

The 0 resonances are expected near the location where the phase
is null, for a wavelength $\lambda > \lambda_p$ but very close 
to $\lambda_p$. When $h$ goes towards zero, the position of the 0 resonance
will come very close to $\lambda_p$. Since there is a zero of $t_0\,t_1$
for $\lambda=\lambda_p$, the 0 resonance is almost non existent in this limit.
Finally, the $0$ resonances cannot produce a very enhanced
transmission under a certain thickness.

A typical spectrum is shown figure \ref{fig8} where a 0
resonance is clearly seen. For $\lambda=\lambda_p$ a zero of the
transmission can be observed. This is what happens
in\cite{garcia-vidal} at low thicknesses. Let us stress that 0
resonances have been widely studied\cite{garcia-vidal,lalanne3} and
are not new.

The geometrical parameters of the structure have been obtained using
a genetic algorithm in order to get the best transmission possible at
$600\,nm$. More precisely, the objective function which is minimized
is $|1-T(600)|+T(400)+T(750)$ where the transmittance $T$ is computed
using the full RCWA (with no approximations). The thickness is
limited to $150\,nm$ in order to have only one resonance and thus a
monochromatic filter. Such resonances are usually observed for
relatively large values of the slit width\cite{porto} and we imposed $a>70\,nm$.
Even if the resonance provides a very intense
transmission, the slits are so wide that the propagative mode is
easily excited. The transmission is quite high at any wavelength
(except near the plasmon resonance). The one-mode approach is
perfectly valid for these conditions.

The $\pi$ resonances are found when the phase is close to
$\pi$. Figure \ref{fig2} shows that this happens near the zero of $r$.
Thus for $a=100\,nm$ no $\pi$ resonance can be excited at all, since 
$|r|\neq 0$ is required to observe a minimum of the denominator 
in (\ref{e:fp}). But figure \ref{fig5} shows that for narrower slits
the zero of $r$ goes off the real axis. Even if $|r|$ is not very
high, this can produce a peak in the transmission which can then be
attributed to a $\pi$ resonance. This peak is generally very narrow
because the phase varies quickly in the range where the $\pi$
resonances appear.

\begin{figure}[htb]
\centerline{\includegraphics[width=8.3cm]{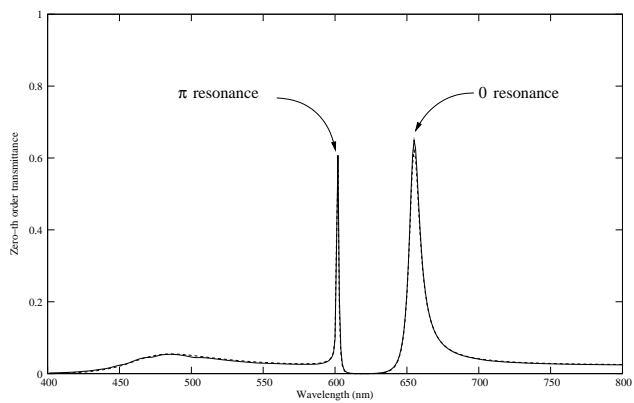}}
\caption{Zero-th order transmission of a slit array with
  $d=600\,nm$,$a=45.5\,nm$ and $h=120\,nm$ (solid line) and the
  single-mode approximation result (dashed line). The $O$ and the
  $\pi$ resonance can be seen on the same spectrum.}
\label{fig9}
\end{figure}

It is possible to observe a $0$ resonance and a $\pi$ resonance on the
same spectrum. Figure \ref{fig9} shows such a situation for
$d=600\,nm$, $a=45.5\,nm$ and $h=120\,nm$. The $\pi$
resonance happens for a wavelength which is around
$602\,nm<\lambda_p$. The $0$ resonance is wider and is found for a
wavelength greater than $\lambda_p$, while there is a zero of
transmission for $\lambda=\lambda_p$. The one-mode model is found to
be very accurate.

\begin{figure}[htb]
\centerline{\includegraphics[width=8.3cm]{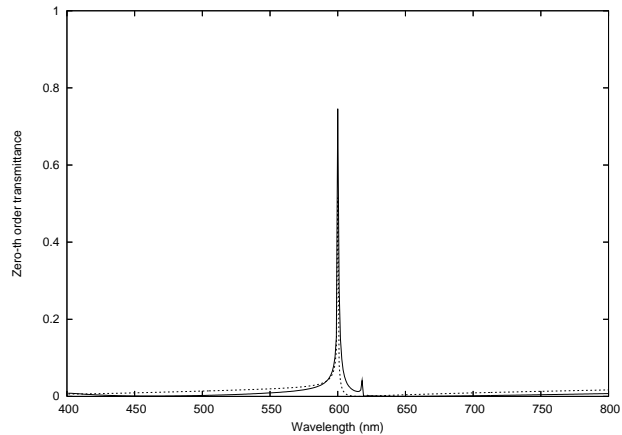}}
\caption{Zero-th order transmission of a slit array with
  $d=600\,nm$,$a=45.5\,nm$ and $h=60.5\,nm$ (solid line) and the
  single-mode approximation result (dashed line). The parameters of
  this structure have been found by optimization, so that the $\pi$
  resonance presents a maximum $@600\,nm$.}
\label{fig10}
\end{figure}

We have performed the same optimization for $\pi$ resonances
than previously, except imposed $a<70$ and a period $d<\lambda_p$
to be sure of the nature of the resonance. The obtained structure has a period of $600\,nm$, a
slit width of $a=45.5\,nm$ and a thickness of only $60.5\,nm$.
The resulting spectrum is shown figure \ref{fig10}. It can be
seen that the one-mode approach is correctly predicting the position,
the width and the intensity of the enhanced transmission even if this
is not the case at other wavelengths. That is why we think that the
peak in the transmission spectrum can be considered as a cavity 
resonance, but with a cavity which is only $60\,nm$ only thick.
The resonance is found at $600\,nm$ precisely. This is
probably the optimal solution because in this case the minimum of the
denominator of (\ref{e:fp}) occurs when there is a maximum of $t_0\,t_1$.

\section{Conclusion}

We have thoroughly studied the behavior of slit arrays for low
thicknesses. We defined horizontal resonances as modes of the
semi-infinite slit array, and vertical modes as resonances of
the finite-thickness structure that cannot be related to horizontal
resonances.

We identified two horizontal resonances. The first one is well known,
it can be found for $\lambda=\lambda_p$ the wavelength for which
surface plasmons are expected. The second one had never been studied
and can be found exactly for $\lambda=d$ provided the slits are wide
enough.

The first one cannot lead to an enhanced transmission. The second one can
lead to an enhanced transmission provided the thickness of the
structure is so small that the single-mode approximation is not
valid any more. Finally, any extraordinary transmission that is accurately described
by the one-mode model is a vertical resonance (or cavity resonance).

We have found a new type of cavity resonance which can be excited for
suprisingly low thicknesses. This type of resonnance can be excited
provided the slits are narrow enough. It can be found around
$\lambda=d$ and always for $\lambda<\lambda_p$. It is a very narrow resonance,
so that slit arrays could constitute very effective monochromatic
filters. We have optimized the transmission spectrum of the structure
in that direction.

We hope this work will provide its readers with a convenient picture of the
optical behavior of extremely thin slit arrays.

\ack

The authors are grateful to Philippe Lalanne for his suggestions and
the attention he paid to our work.

\end{document}